\documentstyle[12pt]{article}
\def\ltap{\raisebox{-.4ex}{\rlap{$\sim$}} \raisebox{.4ex}{$<$}}
\def\gtap{\raisebox{-.4ex}{\rlap{$\sim$}} \raisebox{.4ex}{$>$}}
\def\journal{\topmargin .3in	\oddsidemargin .5in
	\headheight 0pt	\headsep 0pt
	\textwidth 5.625in 
\textheight 8.25in 
	\marginparwidth 1.5in
	\parindent 2em
	\parskip .5ex plus .1ex		\jot = 1.5ex}
\journal
\newskip\humongous \humongous=0pt plus 1000pt minus 1000pt

\newif\ifdtup

\begin{document}
\begin{titlepage}
\begin{center}
June 1994 \hfill    LBL- 35776\\

\vskip .5in

{\large \bf Physics at High Energy Photon Photon  Colliders
}\footnote
{This work was supported by the Director, Office of Energy
Research, Office of High Energy and Nuclear Physics, Division of High
Energy Physics of the U.S. Department of Energy under Contract
DE-AC03-76SF00098.}
\vskip .25in
Presented at the Gamma Gamma Collider Workshop\\
Lawrence Berkeley Laboratory\\
March 27-30, 1994\\
To be published in the proceedings.
\vskip .5in

Michael S. Chanowitz\footnote{email chanowitz@lbl.gov}  \\[.5in]

{\em Theoretical Physics Group\\
     Lawrence Berkeley Laboratory\\
     University of California\\
     Berkeley, California 94720}
\end{center}

\vskip .5in

\begin{abstract}
I review the physics prospects for high energy photon
photon colliders, emphasizing results presented at
the LBL Gamma Gamma Collider Workshop.
Advantages and difficulties are reported for
studies of QCD, the electroweak gauge sector, supersymmetry,
and electroweak symmetry breaking.
\end{abstract}

\end{titlepage}

\newpage
\renewcommand{\thepage}{\arabic{page}}
\setcounter{page}{1}
\noindent {\bf Introduction }

This report  is a brief overview of research that
could be performed at a high energy $\gamma \gamma$ collider.
It is based primarily but not exclusively on 15 talks presented in the three
theoretical physics parallel sessions
at the LBL Gamma Gamma Collider Workshop.
Written versions of these talks are (or should be) included
in these proceedings, as are two excellent survey
talks presented at the workshop by Brodsky\cite{brodskyreview}
and Ginzburg\cite{ginzburgintro}.

The ability to obtain $\gamma \gamma$ and $e \gamma$
collisions by back-scattering low energy laser photons from
high energy $e^\pm$ beams\cite{telnov} can significantly enhance
the physics program of  a linear electron
positron collider. With  $\gamma \gamma$ collision energy of
$\simeq~80\%$ of the parent $e^+e^-$ collider and
comparable luminosity, a PLC (photon linear collider) would provide
unique capabilities in addition to some  welcome redundancy.
Measurement of the
two photon decay width of the Higgs boson would alone be
sufficient  motivation to add the $\gamma \gamma$ collision
option to an $e^+e^-$ collider.

Since the workshop is an ecumenical gathering of accelerator
and laser physicists as well as experimental and theoretical
particle physicists, I will preface this report with a few remarks
on the  current status of high energy physics, to establish  the
context within which a $\gamma \gamma$ collider must be viewed.
The starting point is the standard model, which offers a compact
and remarkably successful description of all extent experimental
data.
But the standard model is far from being a complete description
of nature. To list just a few of the open questions, the standard model
\begin{itemize}
\item contains 17 arbitrary, unexplained parameters,
\item unifies the weak and electromagnetic forces,
but leaves unresolved the possibility of the further unification of the strong
and gravitational forces,
\item offers little insight into its own gross architectural structure ---
such as the $SU(3) \times SU(2) \times U(1)$ gauge symmetry and the
number of quark-lepton families,
\item provides a framework (the Higgs mechanism) for mass
generation that implies a new force and associated quanta but
leaves their precise properties unknown...
\end{itemize}

With one exception we are not
sure if, how, or when we will find the answers to these questions
nor to others  I have not mentioned.
The single exception, the problem of mass generation,
necessarily has a very strong claim on our
attention. The standard model {\em predicts} the existence
of a fifth force and associated quanta that give
mass to the quarks, leptons, and massive gauge bosons ($W$
and $Z$).  To account for the masses
of the $W$ and $Z$ bosons, the new force must begin to emerge
at an energy scale no greater than about 2 TeV.\cite{mcannrev} This is a
landmark in what is otherwise an unmarked wilderness.
(The next unequivocal landmark is
the Planck mass, at $10^{19}$ GeV, a scale not likely to fall within the
purview of accelerator physics for the next few millenia.)

The prediction of a fifth force
follows from the Higgs  mechanism, which
is an essential feature of the standard model. Like any
prediction in science, this prediction could fail. If it fails
the standard model fails. But the TeV scale landmark still
stands, since we would then discover a deeper theory that
has masqueraded until now as the standard model. The effects of the new
theory would begin to emerge in the same energy region in which the
fifth force must emerge if the standard model is correct.

We are all going to be very surprised if the Higgs mechanism
fails to explain the $W$ and $Z$ boson masses.
But outside particle physics it is not widely understood that
the Higgs {\em mechanism} does not necessarily imply the
existence of Higgs {\it bosons}. There are actually two
possibilities:\cite{mcannrev}
\begin{enumerate}
\item The fifth force is weak in which case
 there are Higgs bosons below
1 TeV and perturbation theory can be applied to Higgs sector
interactions.
\item The fifth force is strong in which case we do
not expect Higgs bosons but a more complex spectrum
of strongly interacting quanta, probably
beginning between 1 and 3 TeV, and perturbation theory is inapplicable.
The unequivocal signal for this case is the existence of strong
$WW$ scattering above 1 TeV.
\end{enumerate}
There is a prejudice among many theorists in favor of
 supersymmetry, which would imply a weak fifth force and
at least one
light Higgs boson, with mass $\ltap \ 140$ GeV. But the evidence is
far from definitive and we should prepare for either possibility.

The LHC operating at its 14 TeV design energy and its
$10^{34}{\rm cm}^{-2}\ {\rm sec}^{-1}$ design luminosity will probably be
able to determine the strength of the fifth force whether weak or strong
and to
provide the first glimpses of the associated new quanta.\cite{mcwk,jbetal}
To have the same capability an $e^+e^-$ linear collider would
need center of mass energy of at least $\sim$ 2 TeV and luminosity
$\sim 10^{34}$ cm$^{-2}$ sec$^{-1}$,\cite{najimakurihara,hkm,mckek}
which will not be possible until well after the expected start date of the
LHC. But whatever is glimpsed at the LHC will not be understood
without  exhaustive further study, at which an
$e^+e^-$ linear collider should excel. To evaluate the physics
potential of an $e^+e^-/e\gamma/\gamma\gamma$ linear
collider complex, we focus on its analyzing power more than
simply on its discovery potential. The LHC should tell us a great
deal about the  energy and luminosity a linear collider would need for
detailed studies of the symmetry breaking sector.
Today, in our ignorance, we must consider a range of
possibilities.

In the following sections I will review the theoretical contributions
to the workshop as well as some other relevant material. Topics
include QCD, the electroweak gauge sector, supersymmetry, and electroweak
symmetry breaking in both the weak and strong fifth force scenario.
In view of the preceding remarks it will come as no surprise that
nearly two thirds of the contributed talks concerned electroweak
symmetry breaking.

\vskip .25in
\noindent{\bf  QCD}

In this section I will sketch two topics in QCD that could be
studied advantageously at an $e^+e^-/e\gamma/\gamma\gamma$
collider complex: the photon structure functions and the top quark threshold
region.

\noindent{\it \underline{ Photon Structure Functions}}

This is a subject that the $e\gamma$ collider {\em owns}.
The inclusive scattering process
$$
e + \gamma \rightarrow e/\nu + X\;,
$$
where $X$ represents any hadronic final state, is
mediated by exchange of a highly virtual $\gamma$, $Z$, or $W$, and
probes the short distance hadronic structure of the photon, just
as deep inelastic electron nucleon scattering probes internal
nucleon structure. Deep inelastic scattering from a photon target
has some unique properties: the structure function $F_2$
increases logarithmically with the four momentum $Q^2$ of the virtual
exchanged gauge boson {\em and} is completely determined in the $Q \to \infty$
limit by perturbative analysis,\cite{witten} in which limit it dominates
the cross section by virtue of the logarithmic enhancement.
This contrasts with the nucleon structure
functions, for which the scaling laws (and
their QCD corrections) are predicted but the functional form
cannot be determined perturbatively.

Because of the experimental difficulty of
isolating the leading photon structure
function, the predicted scaling law and functional form have not
been definitively tested. A high energy $e\gamma$ collider would
offer the best chances to carry out these fundamental measurements.
I am not aware of  feasibility studies for such a program.
It is clearly  worth studying.

At the workshop Frances Halzen presented  a very nice talk
outlining a method to extract the gluonic component of
the photon structure function.\cite{halzentalk}
The gluonic component is not determined by perturbative
analysis and is important for a variety of applications, including
background estimates for $\gamma\gamma$ collisions and in
cosmic ray physics. The idea is to measure the rapidity distribution
for production of heavy quark pairs, $\overline bb$ or $\overline cc$.
Halzen and collaborators Eboli and Gonzalez-Garcia
observe that the signal in
the extreme backward direction (the target fragmentation
region) is overwhelmingly dominated by the gluonic component
of the target photon structure function. Measurement of the
$\overline bb$ or $\overline cc$ cross sections in this region then
provides a measurement of the gluonic component.

The observation is made plausible by the
fact that it holds for a wide range of model structure functions.
However its generality is not clear to me
nor how it might be tested. Since the analysis was ``fresh off the
blackboard'' at the time of the workshop, these issues may
be addressed in the future.

\noindent{\it \underline {Top quark threshold region}}

This subject was not studied at the workshop
but since it is potentially very interesting I will briefly review it.
There are tantalizing possibilities to
study the $\overline tt$ threshold region at a $\gamma
\gamma$ collider, though it remains to be seen how well they can actually be
implemented.

For experimentally relevant masses, $m_t > 150$ GeV, the top
quark lifetime is shorter than the characteristic time scale of
strong interactions (i.e., $\Gamma_t > \Lambda_{QCD}$), so that
the top quark decay $t \rightarrow bW$ occurs before toponium
formation can occur.  Therefore we do not expect narrow toponium
resonances like the charmonium and bottomonium states that taught
us so much about QCD. That was  the bad news. The good news,
heralded by Fadin and Khoze, is that the broad top quark decay width
provides an infra-red cutoff so that the entire threshold region
can be studied with perturbation theory.\cite{fadinkhoze}
The running coupling constant is evaluated at the scale
$$
\alpha_S = \alpha_S(m_t \sqrt{\Gamma_t^2 + E^2})
$$
where $E = \sqrt{s} -2m_t$, and therefore never becomes
nonperturbatively large.

There are then some interesting possibilities:
\begin{itemize}
\item The shape and position of the
$\gamma \gamma \rightarrow \overline tt$ threshold
enhancement determine $m_t$ and $\alpha_S$,
though the beam energy spread dilutes the quality of the
measurement.\cite{ggtt,bigietalpwave}
\item With $\geq 95\%$ polarized photon beams of opposite helicity,
$\lambda_1 \lambda_2 = -1$, which
suppresses the dominant $s$-wave,
production of
$\overline tt$ in the $p$-wave could be observed,\cite{bigietalpwave} with
possible precise determinations of $\alpha_S$ and $m_t$.
In $e^+e^-$ collisions the $s$-wave cannot be similarly suppressed but it may
still be possible to probe the $p$-wave by measuring its interference with the
$s$-wave.\cite{eettpwave}
\item We could measure the important and inaccessible
top quark decay width {\em if} we could obtain energy resolution
$\Delta E_{\gamma \gamma} \ \ltap \ 1$ GeV.  For now this seems
like asking for a
perpetual motion machine, since the only known way to decrease the energy
spread is by increasing the distance between the conversion point and the
interaction point, with a loss of luminosity proportional to the square of the
energy spread.
\item With linearly polarized photon beams we could measure $t$ quark
polarization induced by QCD final state interactions, providing a precise
determination of $\alpha_S$,
and probe for interactions outside the standard
model.\cite{fadinetalpoln,bernreutheretal} These
polarization effects are expected to survive the energy spread of the beams.
\end{itemize}

Time will tell how practicable these proposals are.

\vskip .25in
\noindent{\bf Electroweak Gauge Sector}

Photon photon scattering is the process of choice for testing the
interactions of the electroweak gauge sector, since we begin with
two gauge bosons in the initial state.  It is not surprising that it affords
the most sensitive probes of gauge sector interactions for a given
$e^+e^-$ collider energy.

The dominant process  is
$\gamma \gamma \rightarrow WW$, which has a large, asymptotically
constant cross section,
$$
\sigma = {8\pi\alpha^2 \over M_W^2} \sim 93\ {\rm pb},
$$
corresponding to $\sim 10^6$ $W^+W^-$ pairs per 10 fb$^{-1}$.
Compared to the  point-like photon mediated cross section
$\sigma_{POINT}(e^+e^- \rightarrow \mu^+ \mu^-)$, the traditional ratio
$R$ grows with energy,
 $$
R(\gamma \gamma \rightarrow WW) = {\sigma(\gamma \gamma \rightarrow WW) \over
\sigma_{POINT}} = { 6s \over M_W^2}
$$
where $s$ is the square of the total center of mass energy.
Other $ 2 \to 2$ processes in $\gamma \gamma$ scattering
and $e^+e^-$ annihilation have cross sections that fall like $s^{-1}$
(up to logarithms in some cases). At $\sqrt{s}=500$ GeV we have
$R(\gamma \gamma \rightarrow WW) \sim 230$, an order of magnitude
larger than $R(e^+e^- \rightarrow WW) \sim 18$ at the same energy.

This is another instance of the  particle physics  maxim ``yesterday's Nobel
prize, tomorrow's background.''
The large $WW$ cross section is advantageous in testing for
anomalous gauge sector interactions but is a decided disadvantage
in many searches for new physics for which it provides an enormous
background. This will be evident  in the discussions of
Higgs sector and supersymmetry signals in the next sections.

The $WW$ cross section is not as overwhelming as the
above equations seem to suggest. The constant total cross section
arises from singularities in the forward and backward directions, and
as the energy increases the scattering becomes more and more
concentrated at small scattering angles. The cross section for
scattering greater than a fixed angle $\theta > \theta_0$ has the
conventional scaling behavior,  falling
like $s^{-1}$. Integrating over all angles we have
schematically
$$
\sigma \sim \int dt{1\over(t-M_W^2)^2} \sim {1\over M_W^2}
$$
whereas at large $s$ with $\theta > \theta_0$
$$
\sigma \sim \int_{\theta > \theta_0} dt{1\over(t-M_W^2)^2}
 \sim {1\over s}{1 \over (1-{\rm cos}\theta_0)}.
$$

The effect of the scattering angle cut is shown in table 1 for
$\gamma \gamma$ collisions at 0.5, 1.0, and 2.0 TeV. Though it
reduces the cross section tremendously, especially at the highest
energies, the surviving cross sections are still very big relative
to typical signal cross sections of interest. In practice it is
not possible to cut on the center of mass scattering angle because
of the energy spread of the photon beams. In a study
of supersymmetry signals
described in the next section,  Murayama and Kilgore\cite{kilgoremurayama}
find that it is more effective to cut on the transverse momentum of
the $W$ or its decay products than on the laboratory scattering angle.

Following the principle  ``when you've got lemons make lemonade,''
it is worth considering whether a PLC could be used as a $W$ factory.
Is there an interesting physics program in high statistics
studies of $W$ boson decays? To stimulate consideration of the question and
to provide guidance toward a constructive answer, I announced the
Second Chanowitz Prize\cite{mckek}
at the Second KEK Topical Conference on $e^+e^-$ Collisions: lunch with
Michael Peskin for suggesting an interesting $W$ factory program
( Chez Panisse in Berkeley) or for proving a no-go theorem ( SLAC
cafeteria). As of this writing the prize is still unclaimed.

As shown first by Jikia\cite{jikiaggzz} and confirmed
analytically\cite{bergerggzz} and numerically\cite{numggzz},
the large cross section for $\gamma \gamma
\rightarrow WW$ engenders a surprisingly large cross section for
$\gamma \gamma \to ZZ$ via  the $WW$ intermediate state. Measurement
of $\sigma(\gamma \gamma \to ZZ)$ will be a significant test of
the electroweak gauge sector at the quantum loop level.
Though also sharply peaked in the forward direction,
$\gamma \gamma \to ZZ$ is still a formidable background. Even
after cuts on the scattering angle or transverse momentum, it overwhelms
the Higgs boson signal for $m_H\ \gtap \ 400$ GeV
and obscures the growing contribution to the cross section from
ultraheavy charged quanta\cite{mccount}.

More recently Jikia and collaborators have computed the cross sections
for $\gamma \gamma \to \gamma Z$ \cite{gggz} and
$\gamma \gamma \to \gamma \gamma$,\cite{gggg} which are also
dominated by the $W$ loop contribution. It is  splendid
to imagine measuring the elastic, on-shell scattering
of light by light! With a PLC at a 500 GeV
$e^+e^-$ collider, there would be $\sim 50$ events with scattering
angle $|\theta | > 30^o$ per 10 fb$^{-1}$ of $\gamma \gamma$
luminosity.

Given the two gauge boson initial state, a $\gamma \gamma$ collider
is clearly the premier facility for testing the electroweak
gauge sector interactions of the standard model.  The generic sensitivity
of the three beam combinations at given $e^+e^-$ collider energy is
$\gamma \gamma > e \gamma > e^+e^-$.  This ordering does
not apply to every possible anomalous interaction. For instance,
Eboli and Han presented studies
of $\gamma ZWW$ interactions for which $e \gamma$
collisions have the greatest sensitivity.
Eboli and collaborators\cite{ebolianom} assume
an interaction invariant under $U(1)_{\rm EM}$, $C$, $P$, and
$SU(2)_{\rm Custodial}$ but not under the complete local
$SU(2)_{\rm L} \times U(1)_{\rm Y}$,
$$
{\pi \alpha \over 4\Lambda^2} a_n  W_{\alpha}\cdot  W_{\nu}
\times W_{\mu}^{ \alpha} F^{\mu \nu}\;.
$$
They find for $\Lambda = M_W$ that a 3$\sigma$ constraint
$ -1.2 < a_n < 0.74$ can be achieved with 10 fb$^{-1}$ at a 500 GeV
$e^+e^-$ collider.

Han and collaborators\cite{hananom}
considered a $\gamma ZWW$ interaction that is
locally $SU(2)_{\rm L} \times U(1)_{\rm Y}$
and $CP$ invariant but violates $C$, $P$, and $SU(2)_{\rm Custodial}$,
$$
\hat{\alpha}{\left ( 2e^4 \over {\rm cos}\theta_W
{\rm sin}^3 \theta_W \right ) }
{v^2 \over \Lambda^2} \epsilon^{\alpha \beta \mu \nu}
W^-_{\alpha} W^+_{\beta} Z_{\mu} A_{\nu}\;.
$$
With 10 fb$^{-1}$ at parent $e^+e^-$ colliders of 0.5 and 2.0 TeV they
find 3 $\sigma$ limits of $\hat{\alpha}\
\ltap\ 12$ and $\hat{\alpha} \ \ltap\  1$
respectively for $\Lambda = 2$ TeV.
The results are very sensitive to the scattering energy, much less
sensitive to the luminosity.

In some cases enhanced sensitivity can be achieved
by combining data from all three beam combinations of an
$e^+e^-/e \gamma/ \gamma \gamma $ collider. This was nicely illustrated
by Choi and Schrempp\cite{choischrempp}, who showed
that the constraint on the anomalous magnetic moment of the
$W$ obtained at
a 500 GeV collider is vastly improved by combining measurements
from all three collision options.

\vskip .25in
\noindent{\bf Supersymmetry}

Murayama presented the results of a study prepared
for the workshop in collaboration with Kilgore,
to compare the scalar muon signal at
a $\gamma \gamma$ collider with the signal at an $e^+e^-$
collider.\cite{kilgoremurayama}
The emphasis is not simply on discovery potential but on the ability
to make a precise measurement of the mass. If supersymmetry is
discovered such measurements will be extremely important since
they would then test theories at much higher energy scales, such
as supergravity, for which the natural scale is only a few orders of magnitude
below the Planck mass. The scalar muon is also a prototype for many
other measurements and searches
that use lepton and missing energy signals and
are therefore vulnerable to a large $WW$ background.

The signal is $\gamma \gamma \to {\tilde \mu}^+ {\tilde \mu}^- \to \mu^+ \mu^-
 + {\rm LSP\ \overline{LSP} }$ where LSP refers to the lightest
supersymmetric particle, which escapes from the detector like
a neutrino. A dangerous background is then $\gamma \gamma \to
W^+W^- \to \mu^+ \mu^- + \overline \nu \nu$. The signal is enhanced
by a factor $\sim 2$ relative to the background
by choosing photon beams of equal helicity so that $J_Z = 0$, but
before additional cuts the surviving background is still at least
10 times larger than the signal. Assuming a 150 GeV smuon with
$m_{\rm LSP}=100$ GeV and a 500 GeV $e^+e^-$ collider,
Murayama and Kilgore
 eliminate the background by an acoplanarity cut and a cut
on the muon transverse momentum. The surviving, essentially pure
signal has a 20 fb cross section, so 10 fb$^{-1}$ is more than adequate
for discovery.

It is necessary to cut hard enough to obtain an essentially
pure signal sample in order to make an accurate measurement of
the smuon mass. With 50 fb$^{-1}$  a 5 GeV measurement
of the mass is possible.\cite{kilgoremurayama} While
impressive this does not match the 1 GeV accuracy that
can be obtained from 500 GeV $e^+e^-$ collisions with 20 fb$^{-1}$
using right hand polarized electrons to remove the $WW$
background.\cite{tsukamotoetal}
The increased accuracy is due in part to the smaller energy spread of
the $e^+e^-$ beams. Increasing $\rho$
(the distance from the $e \gamma_{\rm Laser}$ conversion point
to the $\gamma \gamma$ interaction point)
decreases the $\gamma \gamma$
energy spread but at too great a cost in luminosity.

This study indicates the generic difficulty of using $\gamma \gamma$
collisions for such measurements, due to
the large $WW$ background and the large spread in photon energies.
At higher energy colliders beamstrahlung also spreads the
the $e^+ e^-$ center of mass energy,
reducing  the relative advantage of $e^+e^-$ collisions.

As mentioned by Murayama,
a $\gamma \gamma$ collider has a great advantage over its parent
$e^+e^-$ collider for the study of heavy scalar superpartners
such as the top squark or stop, $\tilde t$.
In $e^+e^-$ collisions stop-antistop
would be produced in the kinematically suppressed $p$-wave and
could not be effectively studied unless the available collider energy
were much greater than the threshold production energy. In
$\gamma \gamma$ collisions stop-antistop pairs are produced in the $s$-wave,
which can be further enhanced by choosing photon beams of equal helicity.

\vskip .25in
\noindent{\bf Electroweak Symmetry Breaking}

Though more careful studies are needed to be sure, it is likely that
the LHC at design energy and luminosity can provide observable signals
of the strong $WW$ scattering that occurs at $\sqrt{s_{WW}}\ > \ 1$ TeV
if the symmetry breaking fifth force
is strong.\cite{mcwk,jbetal} Those measurements determine the energy
scale of the fifth force and associated quanta whether they detect a signal or
not, since the absence of strong
scattering signal would imply  a weak fifth force and Higgs bosons below
$\simeq 1$ TeV. Higgs bosons themselves
would also be observable at the LHC,
though with difficulty in the ``intermediate mass region'' below the
$ZZ$ threshold and above the $\sim 80$ GeV reach of LEP II. The
supersymmetric Higgs bosons are more difficult to observe at LHC than
the Weinberg-Salam Higgs boson, but supersymmetry itself is likely
to be easily discovered since the strongly interacting superparticles
(squarks and gluinos) would be produced with sizeable cross sections.

Higgs bosons are readily observable at $e^+e^-$ colliders
given sufficient energy and luminosity. To cover the
mass range from the current 60 GeV limit to the likely upper limit
of $\sim 1$ TeV, we would need  a collider with total
energy $\sqrt{s} \geq {\rm MIN}( m_H + m_Z,\ m_H/0.7)$
and integrated luminosity ranging from 1 fb$^{-1}$ at the low end
to $\simeq 200$ fb$^{-1}$ at the upper
end.\cite{najimakurihara,hkm,mckek} The Higgs bosons of
supersymmetric theories are more readily observable at
$e^+e^-$ colliders than at hadron colliders.

The question  then is ``What does a $\gamma \gamma$ collider bring
to the party?'' There are, at least, the following answers:
\begin{itemize}
\item ability to measure $\Gamma(H\to \gamma \gamma)$ for
$m_H\ \ltap\ 350$ GeV --- a fundamental measurement as described
below,
\item extending the reach of an
$e^+e^-$ collider for the most elusive supersymmetric Higgs bosons, the
heavy scalar $H^0$ and the pseudoscalar $A^0$,\cite{gunionhaber}
\item complementary observations of the charged Higgs bosons $H^\pm$
of nonminimal Higgs sectors\cite{hpm},
\item circular and linear polarization of the photon beams offer
unique analyzing power,e.g., to measure the parity of the Higgs
bosons\cite{gunion,krameretal} and to enhance signals relative to backgrounds,
\item ability to observe strong $WW$ scattering in $\gamma \gamma
\to WWWW, WWZZ$\cite{brodskywaikaloa,jikiastrong,cheungstrong}
and to observe strong
$WW$ resonances in $\gamma \gamma \to
ZZ$,\cite{mbmc,otherggzzstrong}
though in colliders of the far future with $\sqrt{s}\ \gtap\ 2$ TeV.
\end{itemize}
These topics are reviewed below.

\noindent{\it \underline {Higgs Bosons}}

A $\gamma \gamma$ collider is the facility of choice to measure
the $\gamma \gamma$ decay width of the Higgs boson. This is not just
an important test of the Higgs theory but also
probes the existence of arbitrarily heavy quanta that may be
far too heavy to produce in existing or even presently contemplated
accelerators.\cite{egn}
The $H \to \gamma \gamma$ decay
proceeds via all intermediate quanta that
are electrically charged and receive mass from the Higgs boson.
All such quanta of spin 0 or 1/2
that are heavier than $m_H$ contribute depending only on
their spin and electric charge but {\em independently of how heavy
they may be.} Consequently $\Gamma(H \to \gamma \gamma)$ is an
amazing window to the highest mass scales that are coupled to the
Higgs sector.

Several presentations at the workshop considered the question of
how to detect a Higgs boson with mass below the $ZZ$ threshold, which
would decay predominantly to $\overline bb$. The problem is how to
see the signal over $\overline bb$ backgrounds from ``direct''
$\gamma \gamma \to \overline bb$ production and from $\overline bb$
production by ``resolved'' photons\cite{resolvedggbkgd} which are produced
predominantly by scattering from the gluon component of the photon.
The resolved photon background is large but soft and can be controlled
by choosing the $e^+e^-$ energy so that the Higgs signal occurs at the
maximum $\gamma \gamma$ energy,\cite{gunionhaber} $m_H \sim
E_{\gamma \gamma}^{\rm MAX} \sim 0.8 E_{e^+e^-}$. Essentially no
$\overline bb$ pairs from resolved photons
occur at the upper edge of phase space, since they
are produced in association with other internal quanta of the
photon.

The leading order direct background is controlled by choosing equal
helicity photon polarizations so that $J_Z =0$, in which case
 $\gamma \gamma \to \overline bb$ is  suppressed
by a factor $m_b^2 / s$ in the cross section.\cite{gunionhaber,bordencaldwell}
 (The suppression follows from the chiral invariance of QCD
interactions which
forbids creation of a massless fermion-antifermion pair with
$J_Z=0$.) As discussed by Borden and
Jikia in presentations at the workshop\cite{borden,jikiabb} the
kinematical suppression does not apply to the leading QCD correction,
$\gamma \gamma \to \overline bbg$,
since after gluon radiation the ${\overline bb}$ system need not be
in a $J=0$ state. Unless it can be controlled the surviving background
would overwhelm the signal. While differing in some respects, both
studies concluded that the background can be controlled with additional
cuts. Borden estimated that a 10\% measurement of the decay width
could be accomplished with 10 -- 20 fb$^{-1}$. A critical requirement is
90 -- 95\% rejection capability for ${\overline cc}$.

Above $ZZ$ threshold $\gamma \gamma \to H \to ZZ$ must be
distinguished from the huge $WW$ background discussed in the previous
section. This rules out the four jet final state, since even
with perfect jet-jet mass resolution intrinsic smearing from
the $Z$ and $W$ widths
may submerge the $ZZ$ signal  in the tail of the $WW$ background.
It is probably necessary to tag at least one of the $Z$'s, either
in its electron or muon decay mode or perhaps in the neutrino mode,
i.e., $ZZ \to l^+ l^- + jj$ with $l=e,\mu$ (net branching ratio from $ZZ$
$\sim 10$\%), or $ZZ \to \overline \nu \nu + jj$ (net branching ratio
$\sim 40$\%). This works for $m_H\ \ltap\ 350$ GeV,
beyond which the signal begins to sink into the $ZZ$ continuum
background.\cite{jikiaggzz,bergerggzz,numggzz}
The width $\Gamma_{\gamma \gamma}$ can be measured to
$\sim 10\%$ at the lower end of the $ZZ$ mass range (more precisely,
$\Gamma_{\gamma \gamma}B_{ZZ}$) but is of course
poorly measured near the upper end as the signal
disappears.\cite{bordencaldwell}

In an $e^+e^-$ collider the supersymmetric Higgs bosons $H^0$ and
$A^0$ are produced in association, $e^+e^- \to HA$. While
a two-for-one sale seems economical, the cost is the energy to
reach the threshold $\sqrt{s_{e^+e^-}} > m_H + m_A$. The claimed
reach at a 500 GeV collider is  $\sim 200$ GeV in the individual
Higgs boson masses. This can be extended using the $\gamma
\gamma$ collider option where $H$ and $A$ can be  produce
individually. For moderate values of the mixing parameter tan$\beta$,
they can be detected decaying to $\overline bb$. The claimed reach for
a parent $e^+e^-$ collider of 500 GeV, using $\gamma \gamma \to
h,H,A \to \overline bb$ is then
to the theoretical maximum, $\sim 145$ GeV,
for the light scalar $h$, the interval $110 < m_H < 200$ GeV
for $H$, and $100 < m_A < 2m_t$ for $A$.\cite{gunionhaber}
The latter significantly extends the reach relative to
the $e^+e^-$ collision mode.

Linear polarization would enable direct measurement of the Higgs
bosons parities.\cite{gunion,krameretal} The scalars $h$ and $H$ couple
couple to the photon polarization vectors like $\epsilon_1 \cdot
\epsilon_2$ while the pseudoscalar $A$ couples like
$\epsilon_1 \times \epsilon_2 \cdot k$ where $k$ is the photon
three-momentum in the center of mass. Kramer {\it et al.}\cite{krameretal}
observe that
linear polarization of 65\% may be obtained by choosing a lower
energy laser (requiring an increase of 1.7 in the $e^+e^-$ energy to
maintain a fixed $\gamma \gamma$ energy). It appears that
100 -- 200 fb$^{-1}$ may be needed to see the asymmetries above
background.

The leading QCD corrections to $\Gamma (H \to \gamma \gamma)$ were
reported by Najima at the workshop.\cite{najimahgg,hggcorrn} The corrections
are very small for $m_H \ < \ m_t$ but are large, of order 1, for
$m_H\ \gg\ m_t$.

\noindent{\it \underline {Strong $WW$ scattering}}

Berger reported on a study,\cite{mbmc}
carried out for the workshop, of strong
interaction effects in $\gamma \gamma \to Z_L Z_L$, where the
subscript $L$ denotes longitudinal polarization. If electroweak
symmetry breaking is due to a strong fifth force, it would be
reflected in the  $\gamma \gamma \to Z_L Z_L$
cross section, which would then be analogous to the hadronic process
$\gamma \gamma \to \pi^0 \pi^0$. This process has  been explored by
others,\cite{otherggzzstrong} though in most instances without
detailed consideration of the very
large $ZZ$ background. Using methods developed in the study of
strong $WW$ scattering at hadron supercolliders, the study
reported by Berger focused on whether the strong scattering
signal would be visible above the large $ZZ$ background.
The conclusion is that nonresonant effects are probably not
observable but that resonances, analogous to the hadronic tensor meson
$f_2(1270)$, could be observed with 100 fb$^{-1}$ and sufficient
energy to produce the resonance. Such resonances are not likely to
occur below $\sim 2$ TeV.

A more promising method to study nonresonant strong $WW$ scattering
was suggested by Brodsky\cite{brodskywaikaloa} and has
been studied at this workshop by Jikia\cite{jikiastrong} and
Cheung.\cite{cheungstrong}
In analogy to strong $WW$ scattering
at $pp$ colliders\cite{mcmkg}, $qq\to qqW_L W_L$,
Brodsky proposed considering $\gamma \gamma \to WWW_LW_L$ or
$\gamma \gamma \to WWZ_LZ_L$. (The analogous process for $H$ boson
production, $\gamma \gamma \to WWH$, has been studied by
Baillargeon and Boudjema.\cite{bb}) At the workshop
Cheung\cite{cheungstrong} presented
signals (without backgrounds)
for a variety of strong scattering models, using the effective $W$
approximation and the equivalence theorem. Jikia reported a
complete leading order
calculation of the backgrounds,  $\gamma \gamma \to WWWW$ and
$\gamma \gamma \to WWZZ$, requiring in the first case evaluation
of 240 Feynman diagrams\cite{jikiastrong}. (Cheung has
subsequently also evaluated the backgrounds.\cite{cheungbkgd})

With the background evaluated Jikia estimated the
energy and luminosity necessary to observe heavy Higgs bosons and
strong $WW$ scattering. With 200 fb$^{-1}$ he finds that a
$\gamma \gamma$ collider at a
1.5 TeV $e^+e^-$ collider is needed to observe the standard model
Higgs boson with $m_H = 700$ GeV, while a $\gamma \gamma$ collider at a
2  TeV $e^+e^-$ collider
is needed for $m_H = 1$ TeV. From these cases he
concludes that the reach of a
$\gamma \gamma$ collider based on a 2 TeV $e^+e^-$ collider is similar to that
of a 1.5 TeV $e^+e^-$ collider operating in $e^+e^-$ mode assuming equal
$\gamma \gamma$ and $e^+e^-$ luminosities, not surprising since
$\sqrt{s^{\rm MAX}_{\gamma \gamma}} \simeq 0.80 \sqrt{s_{e^+e^-}}$.
He incorporates the effect of experimental efficiencies
by consulting the study of
heavy Higgs boson production and strong $WW$ scattering
in $e^+e^-$ collisions by Kurihara and
Najima\cite{najimakurihara}, who did include detector simulation;
they found  for 2 TeV $e^+e^-$ collisions that
$\sim 300$ fb$^{-1}$ is needed to obtain a 3 $\sigma$ strong $WW$
scattering signal. Jikia then infers that a
$\gamma \gamma$ collider at a 2 TeV $e^+e^-$ collider
(with $\sqrt{s^{\rm MAX}_{\gamma \gamma}} \simeq 1.6$ TeV) could not
observe strong $WW$ scattering unless $\gamma \gamma$ luminosities much larger
than $O(200)$ fb$^{-1}$ are possible.
Without attempting to incorporate detector simulation,
Cheung\cite{cheungbkgd} concludes more optimistically
that strong scattering could be seen with $\sim$ 100 fb$^{-1}$ with a
$\gamma \gamma$ collider at a 2 TeV $e^+e^-$ collider, and that
$\sim 10$ fb$^{-1}$ could suffice at a 2.5 TeV $e^+e^-$
collider.

\vskip .25in
\noindent{\bf Conclusion}

The presentations at this workshop show that a photon linear collider
would be a valuable adjunct to the $e^+e^-$ linear collider on
which it would be based. Relative to the parent  $e^+e^-$ collider, the
$\gamma \gamma$ collider suffers from proportionately larger $WW$
backgrounds and, especially in the NLC energy range, from broader
beam energy spread. But it provides a variety of significant advantages,
with unique access to some fundamental physics,
using beams that can be customized for different physics
goals.

By  choosing the
relative helicities of the lepton and laser-photon beams, ``broad'' or
``narrow'' band beams can be provided, with the narrow band beam offering
much of its luminosity at the highest energies, typically
$\sim 80\%$ of the parent $e^+e^-$ collider energy. Increasing the distance
between the conversion point and the interaction point improves the
monochromaticity further, though at a cost in luminosity proportional
to the square of the decrease in energy spread. Circular polarization
is readily achieved and enhanced linear polarization is possible
by lowering the energy of the laser photons.

There is a range of studies for which $\gamma \gamma$ and $e \gamma$
colliders would
be uniquely suited. The $e\gamma$ collider mode is the facility of choice
for probing the photon structure functions, a fundamental
subject in QCD with important phenomenological implications.
Valuable measurements of the $\overline tt$ threshold region may be
possible in $\gamma \gamma$ collisions,
especially with polarized beams. If supersymmetric
particles exist at the electroweak scale, a
$\gamma \gamma$ collider  would be optimal for detailed study
of heavy squark-antisquark states that are suppressed by $p$-wave phase
space in $e^+e^-$ collisions. Since all the initial energy is concentrated in
two gauge bosons, $\gamma \gamma$ collisions offer the most sensitive
probes of the electroweak gauge sector for given $e^+e^-$ collider energy.
For $m_H\ \ltap \ 350$ GeV the $\gamma \gamma$ collider provides the
best (and for $m_H\ > \ 2m_W$ probably the only) measurement of the
two photon decay width of the Higgs boson. It can extend the reach of the
parent $e^+e^-$ collider for the pseudoscalar and heavy scalar Higgs bosons
of supersymmetric models. In addition to its unique capabilities, a
$\gamma \gamma$ collider would provide welcome redundancy
with measurements from $e^+e^-$ and proton-proton collisions.

Aided by my nearly perfect ignorance of accelerator physics and of
linear colliders
in particular, I can imagine another way in which high energy $\gamma
\gamma$ colliders could be crucial. Though unlikely, it is possible
that the ratio of luminosities
${\cal L}_{\gamma \gamma} / {\cal L}_{e^+e^-}$ might be large not just
by virtue of an enhanced numerator, as discussed at this workshop, but
also if the denominator is unexpectedly small. The issues that determine
the luminosity of $e^+e^-$ collisions are not identical with those that
determine the $\gamma \gamma$ luminosity, and unanticipated difficulties
might affect one but not the other. If for instance
unexpected beam-beam instabilities were found to suppress TeV $e^+e^-$
luminosities below the necessary
10$^{33}$ to 10$^{34}$ cm$^{-2}$ sec$^{-1}$ level, it might still be possible
to obtain the necessary luminosities in $\gamma \gamma$ and $e \gamma$
collisions. The $\gamma \gamma$ collider would then be the only game
in town, and its ``redundant'' access to many subjects I have
not discussed would become crucial.

Though we are
still at an early stage in our thinking --- about both the accelerator
and particle physics --- it seems clear that an
$e^+e^-/e\gamma/\gamma \gamma$ collider complex would be a
very useful extension of a linear $e^+e^-$ collider. Continued R\&D is
surely a prudent investment.

\newpage
\noindent {\bf Table 1:} Cross section in picobarns
for $\gamma\gamma \longrightarrow
W^+W^-$ for various $\gamma \gamma$ center of mass
energies and minimum scattering angles.

\begin{center}
\vskip 20pt
\begin{tabular}{c|c|c|c}
$\sqrt{s}$(TeV) & $\sigma_{\rm total}$ &$\cos \theta < 0.8$&
$\cos \theta < 0.6$\cr
\hline
\hline
 0.5 &77 &9.7 &3.1\cr
\hline
1.0 &88 &2.9 &0.86  \cr
\hline
2.0& 91  &0.78 &0.22

\end{tabular}
\end{center}

\end{document}